# Algorithm and Circuit of Nesting Doubled Qubits

Artyom M. Grigoryan*[1] and Sos S. Agaian [2]

[1] Department of Electrical and Computer Engineering,
The University of Texas at San Antonio, USA
<u>amgrigoryan@utsa.edu</u>

[2] City University of New York / CSI

**ABSTRACT**

Copying the quantum states is contradictory to classical information processing since the fundamental difference between classical and quantum information is that while classical information can be copied perfectly, quantum information cannot. However, this statement does not rule out the risk of building a device that can reproduce a set of quantum states. This paper investigates the naturally arising question of how well or under what conditions one can copy and measure an arbitrary quantum superposition of states. The CNOT and XOR operation-based quantum circuit is presented that exhibits entanglement of states and allows for measuring the doubled qubits.

## 1. INTRODUCTION

In 1982, Wootters and Zurek showed that no unitary process can create exact copies of arbitrary quantum states [1]. The statement that non-orthogonal quantum states cannot be copied was also made by Dieks [2], and new no-cloning theorems can be found in [3]. Many publications exist devoted to the no-cloning theorem, that use the linearity and unitarity of the copying transformations [4-6], perfect cloning of no commuting mixed states [7], perfect cloning with assistance [8], multiple copying of qubits [9], supplementary information [10], additional measurement for probabilistic quantum cloning [11,12], cloning with local operation and classical communication [13]. With such an assertion, the theory of quantum computing has become particularly distinctive from computing on traditional computers. In addition, such theorems were used to justify the security of quantum cryptography [14,15].

This paper presents our vision of transferring and copying qubits. In Section 2, we discuss the above statement and present examples where copying of qubits is possible for known qubits. In general, any two qubits can be transferred into each other. Also, in separate Section 3, we analyze a simple quantum circuit with two CNOT and XOR operations for nesting the doubled qubits in 3- qubit. Such a circuit allows us to observe and measure the duplicated states of a qubit.

## 2. BACKGROUND

In this section, we discuss the known statement about copying qubits. The unitary operator $U$ for performing such a copy of an individual qubit in a superposition

$$|\varphi\rangle = a|0\rangle + b|1\rangle = \begin{bmatrix} a \\ b \end{bmatrix}$$

is considered to be the 2-qubit operator

$$U(|\varphi\rangle|0\rangle) \triangleq U(|\varphi\rangle \otimes |0\rangle) = |\varphi\rangle^2 = |\varphi\rangle|\varphi\rangle \triangleq |\varphi\rangle \otimes |\varphi\rangle. \qquad (1)$$

Here, the amplitudes are such that $a^2 + b^2 = 1$ (or $|a|^2 + |b|^2 = 1$ in the complex case). Also, $|0\rangle$ and $|1\rangle$ denote the quantum computational basis states of the single qubit,

$$|0\rangle = \begin{bmatrix} 1 \\ 0 \end{bmatrix}, \qquad |1\rangle = \begin{bmatrix} 0 \\ 1 \end{bmatrix},$$

and $\otimes$ is the operation of the tensor product, or Kronecker product, of vectors.

When measuring the qubit, by using the Hermitian projection on the basis $|0\rangle$ and $|1\rangle$, the probability of the outcome $|0\rangle$ is $a^2$ and the probability of outcome $|1\rangle$ is $b^2$. The axioms of quantum mechanics demonstrate that after the measurement, a qubit is collapsed to the measured basis state, so that a qubit will be destroyed in this measurement. Thus, we may perhaps state the well-known no-cloning theorem: no quantum procedure exists that can reproduce perfectly an arbitrary quantum state. Therefore, the measurement is an irreversible operation; in contrast, it is pretty easy to copy information, even in a reversible manner in classical computers. It means that using the well-known teleportation protocol, we may create a perfect replica of the original qubit, but this will be at the cost of destroying information encoded in the original qubit [9,14]. Thus, a) if we are only interested in producing imperfect copies, then it is possible to design machines (actually, to find unitary transformations) which can copy quantum states, b) if we do two identical copies, then the quality of these copies depends on the input state; and c) we may formulate the quantum copying problem goal is to produce a copy of the initial qubit, which is as close as possible to the original state, while the output state of the original qubit is minimally disturbed [14]. However, this theorem does not rule out the possibility of building a device that can copy a particular set of orthonormal quantum states [16]. In quantum computing, the CNOT operation does not allow for copying the qubits, as in traditional computing, when copying the bits. In a digital computer, when copying a bit, a new cell is allocated in the computer's memory, the value of the bit is read, and then this value is written to the cell. Such a read-write-out procedure is likely to take place in quantum systems. Undoubtedly, other operators and quantum circuits are needed here. For each state of a qubit, probably somewhere in space, an identical state is reproduced. Maybe it is not in its pure form, but in some kind of entangled state with other qubits. In other words, it is possible that such copies can be nested in other shells of systems with a large number of qubits, from where they can be measured and processed.

**2.1 No Copying qubits**

Let us consider the common calculations in the statement of copying qubits. We consider the qubit in the Hadamard superposition

$$|\varphi\rangle = (|0\rangle - |1\rangle)/\sqrt{2}.$$

If a unitary transform $U$ copies this qubit and it is a linear operator, we obtain two qubits in the following state:

$$U(|\varphi\rangle|0\rangle) = \frac{1}{\sqrt{2}}(U(|0\rangle|0\rangle) - U(|1\rangle|0\rangle)) = \frac{1}{\sqrt{2}}(|0\rangle|0\rangle - |1\rangle|1\rangle) = \frac{1}{\sqrt{2}}(|0,0\rangle - |1,1\rangle).$$

The state of doubled qubits is

$$|\varphi\rangle^2 = |\varphi\rangle|\varphi\rangle = \left(\frac{|0\rangle - |1\rangle}{\sqrt{2}}\right) \otimes \left(\frac{|0\rangle - |1\rangle}{\sqrt{2}}\right) = \frac{1}{2}(|0,0\rangle + |1,1\rangle - |0,1\rangle - |1,0\rangle).$$

Thus, we obtain different 2-qubit quantum superpositions $U(|\varphi\rangle|0\rangle)$ and $|\varphi\rangle|\varphi\rangle$. What we use in the above calculations is the assumption that if an unitary operator copies a single qubit in the superposition $|\varphi\rangle$, then it copies any other superposition $|\psi\rangle = c|0\rangle + d|1\rangle$ of the qubit, i.e.,

$$\underline{U(|\varphi\rangle|0\rangle) = |\varphi\rangle|\varphi\rangle} \rightarrow \underline{U(|\psi\rangle|0\rangle) = |\psi\rangle|\psi\rangle}. \tag{2}$$

We can consider the quantum states $|\varphi\rangle|0\rangle$ and $|\varphi\rangle|\varphi\rangle$ as 4-D vectors,

$$|\varphi\rangle|0\rangle = |\varphi, 0\rangle = a|00\rangle + b|10\rangle = \begin{bmatrix} a \\ 0 \\ b \\ 0 \end{bmatrix} \tag{3}$$

and

$$|\varphi\rangle|\varphi\rangle = |\varphi, \varphi\rangle = a^2|00\rangle + ab(|01\rangle + |10\rangle) + b^2|11\rangle = \begin{bmatrix} a^2 \\ ab \\ ab \\ b^2 \end{bmatrix}. \tag{4}$$

We are looking for a 4×4 unitary matrix $U$, such that

$$U\begin{bmatrix}a\\0\\b\\0\end{bmatrix} = \begin{bmatrix}t_{0,0} & t_{0,1} & t_{0,2} & t_{0,3}\\t_{1,0} & t_{1,1} & t_{1,2} & t_{1,3}\\t_{2,0} & t_{2,1} & t_{2,2} & t_{2,3}\\t_{3,0} & t_{3,1} & t_{3,2} & t_{3,3}\end{bmatrix}\begin{bmatrix}a\\0\\b\\0\end{bmatrix} = \begin{bmatrix}a^2\\ab\\ab\\b^2\end{bmatrix}. \qquad (5)$$

It is clear that the coefficients of such a matrix will be defined by the values of inputs, $a$ and $b$. If the states of the qubit are known, such a unitary transform is not difficult to construct using a several rotation gates. Here, we should remind that exists the concept of fast unitary transforms, which are called discrete signal-induced heap transforms (DsiHT) [17]. These transforms found effective implementation is signal filtration [18], image enhancement [19], and QR-decomposition of real and complex square matrices [20,21].

## 2.2. DsiHT and Qubits

The DsiHT, namely its complete system of basic functions is defined by one or a few given signals, which are called the generators of the transform. Any two vectors $x$ and $y$ can be transformed into each other using two DsiHTs with the generators being these vectors. In the case with one generator, we can consider DsiHT that are defined by a chain of 2×2 rotation gates. For instance, given vector $v = (x, y)$, the transformation of this vector to the vector $(||v||, 0)$ can be described by the matrix of rotation

$$Tv' = \begin{bmatrix}\cos\vartheta & -\sin\vartheta\\\sin\vartheta & \cos\vartheta\end{bmatrix}\begin{bmatrix}x\\y\end{bmatrix} = \begin{bmatrix}\sqrt{x^2+y^2}\\0\end{bmatrix}. \qquad (6)$$

Here, the angle of rotation is calculated by $\vartheta = \vartheta(x,y) = -\arctan(y/x)$, or $\vartheta = \pi/2$ if $x = 0$. In the case when this vector has the length 1, we obtain $Tv' = [1\ 0]'$. Thus, any superposition of the single qubit $|\varphi\rangle = a|0\rangle + b|1\rangle$ can be transferred to the 1st basis state $|0\rangle$,

$$T|\varphi\rangle = T\begin{bmatrix}a\\b\end{bmatrix} = \begin{bmatrix}1\\0\end{bmatrix} = |0\rangle. \qquad (7)$$

This unitary transform is parameterized by the generator, $|\varphi\rangle = (a,b)'$, which is the superposition of states of the qubit. Therefore, we will write $T = T_{a,b}$, or $T = T_{|\varphi\rangle}$, and $|\varphi\rangle = [T_{|\varphi\rangle}]'|0\rangle$. Here, ′ denotes the transposition of the matrix. Thus, any superposition of the qubit can be initiated from the basis state $|0\rangle$ (the same is true for the basis state $|1\rangle$). The qubit in the superposition of states $|\varphi\rangle$ can change its superposition to $|\psi\rangle$, by two rotations

$$|\varphi\rangle \to |\psi\rangle = [T_{|\psi\rangle}]'T_{|\varphi\rangle}|\varphi\rangle. \qquad (8)$$

Two and multi-qubit superpositions can be transferred to each other in the same way. For instance, with three Givens rotations, the DsiHT, $H$, performs the transform of a vector $a = (a_{0,0}, a_{0,1}, a_{1,0}, a_{1,1})$, with norm $\|a\| = 1$, into the unit vector $e = (1,0,0,0)$. This transform is generated by this vector $a$, i.e., $H = H_a$. In other words, the 2-qubit $|\varphi\rangle$ represented by this vector is transferred to the basis state $|0\rangle$. In similar way, DsiHTs allow to transfer the 2-qubit to other basis states, $|1\rangle, |2\rangle$, and $|3\rangle$ (for more details see [18]). Another vector $b = (b_{0,0}, b_{0,1}, b_{1,0}, b_{1,1})$ with norm equal 1 can also be transformed into the unit vector $e$ by the DsiHT, $H_b$, generated by $b$. Therefore, these two vectors representing the 2-qubit superpositions of states

$$|\varphi\rangle = a_{0,0}|00\rangle + a_{0,1}|01\rangle + a_{1,0}|10\rangle + a_{1,1}|11\rangle$$

and

$$|\psi\rangle = b_{0,0}|00\rangle + b_{0,1}|01\rangle + b_{1,0}|10\rangle + b_{1,1}|11\rangle$$

can be transferred to each other by using two DsiHTs as

$$|\varphi\rangle \to |\psi\rangle = [H_b]'H_a|\varphi\rangle \quad \text{and} \quad |\psi\rangle \to |\varphi\rangle = [H_a]'H_b|\psi\rangle. \qquad (9)$$

Below, we consider a few examples of such transformation of vectors (or 2-qubits).

## A. Examples of 2-Qubit Transform

First, we describe an example of transforming two arbitrary 2-qubits, $U: |\varphi\rangle \to |\psi\rangle$.

**Example 1:** Consider two 4-dimentional vectors

$$\boldsymbol{a} = \frac{1}{\sqrt{2}}(1,0,0,1)' \quad \text{and} \quad \boldsymbol{b} = \frac{1}{2}(1,1,1,1)'.$$

These two vectors are not orthogonal. The vector $\boldsymbol{a}$ and $\boldsymbol{b}$ represent the 2-qubit superpositions

$$|\varphi_2\rangle = (|00\rangle + |11\rangle)/\sqrt{2} \quad \text{and} \quad |\psi_2\rangle = (|00\rangle + |01\rangle + |10\rangle + |11\rangle)/2.$$

The matrix of the 4-point DsiHT generated by $\boldsymbol{a}$ is

$$H_a = \frac{1}{\sqrt{2}}\begin{bmatrix} 1 & 0 & 0 & 1 \\ 0 & \sqrt{2} & 0 & 0 \\ 0 & 0 & \sqrt{2} & 0 \\ -1 & 0 & 0 & 1 \end{bmatrix} = \begin{bmatrix} 0.7071 & 0 & 0 & 0.7071 \\ 0 & 1 & 0 & 0 \\ 0 & 0 & 1 & 0 \\ -0.7071 & 0 & 0 & 0.7071 \end{bmatrix} \quad (10)$$

with one rotation by -45°. The matrix of the DsiHT with the generator $\boldsymbol{b}$ is

$$H_b = \frac{1}{2}\begin{bmatrix} 1 & 1 & 1 & 1 \\ -\sqrt{2} & \sqrt{2} & 0 & 0 \\ -0.8165 & -0.8165 & 1.6330 & 0 \\ -0.5774 & -0.5774 & -0.5774 & 1.7321 \end{bmatrix} \quad (11)$$

with three rotation by angles -45°, -35.2644°, and -30°. Thus,

$$H_b = \begin{bmatrix} 0.8660 & 0 & 0 & 0.5 \\ 0 & 1 & 0 & 0 \\ 0 & 0 & 1 & 0 \\ -0.5 & 0 & 0 & 0.8660 \end{bmatrix}\begin{bmatrix} 0.8165 & 0 & 0.5774 & 0 \\ 0 & 1 & 0 & 0 \\ -0.5774 & 0 & 0.8165 & 0 \\ 0 & 0 & 0 & 1 \end{bmatrix}\begin{bmatrix} 0.7071 & 0.7071 & 0 & 0 \\ -0.7071 & 0.7071 & 0 & 0 \\ 0 & 0 & 1 & 0 \\ 0 & 0 & 0 & 1 \end{bmatrix}.$$

The matrices $H_a$ and $H_b$ are unitary, and we can write the following: $([H_b]'H_a)\boldsymbol{a} = \boldsymbol{b}$. Thus, we obtain the unitary matrix

$$\boldsymbol{U} = [H_b]'H_a = \begin{bmatrix} 0.5577 & -0.7071 & -0.4082 & 0.1494 \\ 0.5577 & 0.7071 & -0.4082 & 0.1494 \\ 0.5577 & 0 & 0.8165 & 0.1494 \\ -0.2588 & 0 & 0 & 0.9659 \end{bmatrix} \quad (12)$$

for the required transformation

$$(|\varphi\rangle|0\rangle) \to |\varphi\rangle^2 = \boldsymbol{U}(|\varphi\rangle|0\rangle) = \boldsymbol{U}\frac{1}{\sqrt{2}}\begin{bmatrix}1\\0\\0\\1\end{bmatrix} = \frac{1}{2}\begin{bmatrix}1\\1\\1\\1\end{bmatrix}. \quad (13)$$

The matrix $\boldsymbol{U}$ is composed by four matrices of rotation. Such a matrix is not unique in a sense that other unitary matrices can also be used for the above transformation. To show this, we write the above equation as

$$\boldsymbol{U}\begin{bmatrix}1\\0\\0\\1\end{bmatrix} = \frac{1}{\sqrt{2}}\begin{bmatrix}1\\1\\1\\1\end{bmatrix}. \quad (14)$$

Now, we consider the matrix equation

$$A\begin{bmatrix}1\\0\\0\\1\end{bmatrix} = \frac{1}{\sqrt{2}}\underbrace{\begin{bmatrix}1 & 0 & 1 & 0\\ 0 & 1 & 0 & 1\\ 0 & -1 & 0 & 1\\ 1 & 0 & -1 & 0\end{bmatrix}}\begin{bmatrix}1\\0\\0\\1\end{bmatrix} = \frac{1}{\sqrt{2}}\begin{bmatrix}1\\1\\1\\1\end{bmatrix}. \qquad (15)$$

The matrix $A$ is unitary with $\det A = 1$ and its inverse is

$$A^{-1} = A' = \frac{1}{\sqrt{2}}\begin{bmatrix}1 & 0 & 0 & 1\\ 0 & 1 & -1 & 0\\ 1 & 0 & 0 & -1\\ 0 & 1 & 1 & 0\end{bmatrix}. \qquad (16)$$

When using the operations with the matrix $A$ in quantum circuits, this matrix can be presented by two simple gates (permutations) as follows:

$$A = \begin{bmatrix}1 & 0 & 0 & 0\\ 0 & 0 & 1 & 0\\ 0 & 0 & 0 & 1\\ 0 & 1 & 0 & 0\end{bmatrix}\frac{1}{\sqrt{2}}\begin{bmatrix}1 & 1 & 0 & 0\\ 1 & -1 & 0 & 0\\ 0 & 0 & 1 & 1\\ 0 & 0 & -1 & 1\end{bmatrix}\begin{bmatrix}1 & 0 & 0 & 0\\ 0 & 0 & 1 & 0\\ 0 & 1 & 0 & 0\\ 0 & 0 & 0 & 1\end{bmatrix}. \qquad (17)$$

The core matrix in this representation is the matrix

$$A_4 = \frac{1}{\sqrt{2}}\begin{bmatrix}1 & 1 & 0 & 0\\ 1 & -1 & 0 & 0\\ 0 & 0 & 1 & 1\\ 0 & 0 & -1 & 1\end{bmatrix} = H_2 \oplus A_2, \qquad (18)$$

where $\oplus$ is the operation of the Kronecker sum of matrices and the 2×2 matrices are

$$H_2 = \frac{1}{\sqrt{2}}\begin{bmatrix}1 & 1\\ 1 & -1\end{bmatrix} \quad \text{and} \quad A_2 = \frac{1}{\sqrt{2}}\begin{bmatrix}1 & 1\\ -1 & 1\end{bmatrix} = \begin{bmatrix}1 & 0\\ 0 & -1\end{bmatrix}H_2.$$

Thus, in this example we got another matrix $A$ that can be used instead of matrix $U$ in Eq. 12. This matrix was easy to construct by hand, since the unnormalized amplitudes of the 2-qubits are binary, 0 and 1.

**Example 2**. For the qubit $|\varphi\rangle = (3|0\rangle + 4|1\rangle)/5$, we consider the following two 4-dimentional vectors corresponding to the qubits $|\varphi\rangle|0\rangle$ and $|\varphi\rangle|\varphi\rangle$:

$$x = \frac{1}{5}(3,0,4,0)' \quad \text{and} \quad y = \frac{1}{25}(9,12,12,16)'.$$

The unitary transform composed by two DsiHTs $H_y$ and $H_x$ with four rotations by angles 53.13°, 38.66°, 39.79°, and −53.13° has the matrix

$$U = \begin{bmatrix}0.5159 & -0.80 & 0.0631 & -0.2999\\ 0.6878 & 0.60 & 0.0841 & -0.3998\\ -0.3367 & 0 & 0.8525 & -0.3998\\ 0.3840 & 0 & 0.5120 & 0.7684\end{bmatrix}.$$

$\det U = 1$ and $Ux = y$. Such a transform is not unique. Using the concept of the strong DsiHT, we get the following unitary matrix with five rotations by 53.13°, 90°, 53.13°, 38.66°, 39.79° (for detail see [18]):

$$U = \begin{bmatrix}-0.4240 & -0.3748 & 0.7680 & -0.2999\\ 0.7680 & -0.4998 & 0.0240 & -0.3998\\ 0.2880 & 0.7809 & 0.3840 & -0.3998\\ 0.3840 & 0 & 0.5120 & 0.7684\end{bmatrix}, \quad \det U = 1.$$

## B. Matrix *U* for the Hadamard Superpositions

*Example 3:* We consider the superposition

$$|\varphi\rangle = a|0\rangle + b|1\rangle = (|0\rangle + |1\rangle)/\sqrt{2},$$

i.e., the case when $a = b = 1/\sqrt{2}$. We will consider the transformation of this qubit without DsiHT; the example is simple and can be solved manually. Equation 5 can be written as

$$U \frac{1}{\sqrt{2}} \begin{bmatrix} 1 \\ 0 \\ 1 \\ 0 \end{bmatrix} = \frac{1}{2} \begin{bmatrix} 1 \\ 1 \\ 1 \\ 1 \end{bmatrix}, \quad \text{or} \quad U \begin{bmatrix} 1 \\ 0 \\ 1 \\ 0 \end{bmatrix} = \frac{1}{\sqrt{2}} \begin{bmatrix} 1 \\ 1 \\ 1 \\ 1 \end{bmatrix}. \tag{19}$$

It is not difficult to construct such a matrix. The matrix *U* is

$$U = \frac{1}{\sqrt{2}} \begin{bmatrix} 1 & 0 & 0 & 1 \\ 0 & 1 & 1 & 0 \\ 0 & -1 & 1 & 0 \\ 1 & 0 & 0 & -1 \end{bmatrix} \tag{20}$$

and it can be written with two permutations (1,2,3) and (1,3,2) as

$$U = \begin{bmatrix} 1 & 0 & 0 & 0 \\ 0 & 0 & 1 & 0 \\ 0 & 0 & 0 & 1 \\ 0 & 1 & 0 & 0 \end{bmatrix} \frac{1}{\sqrt{2}} \begin{bmatrix} 1 & 1 & 0 & 0 \\ 1 & -1 & 0 & 0 \\ 0 & 0 & 1 & 1 \\ 0 & 0 & -1 & 1 \end{bmatrix} \begin{bmatrix} 1 & 0 & 0 & 0 \\ 0 & 0 & 0 & 1 \\ 0 & 1 & 0 & 0 \\ 0 & 0 & 1 & 0 \end{bmatrix}. \tag{21}$$

One can see that this matrix can be obtained from matrix *A* in Eq. 17, by using the CNOT gate $P_1$,

$$U = AP_1 = \frac{1}{\sqrt{2}} \begin{bmatrix} 1 & 0 & 1 & 0 \\ 0 & 1 & 0 & 1 \\ 0 & -1 & 0 & 1 \\ 1 & 0 & -1 & 0 \end{bmatrix} \begin{bmatrix} 1 & 0 & 0 & 0 \\ 0 & 1 & 0 & 0 \\ 0 & 0 & 0 & 1 \\ 0 & 0 & 1 & 0 \end{bmatrix} = \frac{1}{\sqrt{2}} \begin{bmatrix} 1 & 0 & 0 & 1 \\ 0 & 1 & 1 & 0 \\ 0 & -1 & 1 & 0 \\ 1 & 0 & 0 & -1 \end{bmatrix}.$$

Now, we consider the second Hadamard superposition

$$|\varphi\rangle = a|0\rangle + b|1\rangle = (|0\rangle - |1\rangle)/\sqrt{2},$$

i.e., the case when $b = -a = -1/\sqrt{2}$. Then,

$$|\varphi\rangle|0\rangle = a|00\rangle + b|10\rangle = \frac{|00\rangle - |10\rangle}{\sqrt{2}} = \frac{1}{\sqrt{2}} \begin{bmatrix} 1 \\ 0 \\ -1 \\ 0 \end{bmatrix}$$

and

$$|\varphi\rangle^2 = |\varphi\rangle|\varphi\rangle = \frac{1}{2}(|00\rangle - (|01\rangle + |10\rangle) + |11\rangle) = \frac{1}{2} \begin{bmatrix} 1 \\ -1 \\ -1 \\ 1 \end{bmatrix}.$$

The same unitary transform can be used for copying this qubit, i.e.,

$$U \begin{bmatrix} 1 \\ 0 \\ -1 \\ 0 \end{bmatrix} = \frac{1}{\sqrt{2}} \begin{bmatrix} 1 \\ -1 \\ -1 \\ 1 \end{bmatrix}. \tag{22}$$

Thus, for two qubits in the superpositions of states

$$|\varphi_{1,2}\rangle = \frac{|0\rangle \pm |1\rangle}{\sqrt{2}},$$

we use one copying unitary transformation with the matrix $U$. These two states are orthogonal. It is easy to see that only these two qubits $|\varphi_1\rangle$ and $|\varphi_2\rangle$ can be copied by this transformation.

As mentioned above (see Eq. 9), arbitrary quantum superpositions of states can be transformed into another. Does this mean that the quantum superpositions of only known single qubits can be copied? We will answer this question, if we can, in the next work, but for now we leave it to the reader. Unitary transformations for copying qubits are parameterized by amplitudes of quantum superpositions of states of the qubits. They are not universal, i.e., cannot be used to copy any qubits. And it should be so; everything is like in living systems. In order to reproduce its copy, a living object must have its own progenitor.

## 3. Quantum circuit for nesting doubled qubits

In this section, we present a method of nesting the doubled qubits in a larger state. In other words, we discuss the circuit which might be used to calculate the doubled qubits in arms of 3-qubit. The measurement and separation of states of doubled qubits are described.

Let us consider a qubit in the state $|\varphi\rangle = a|0\rangle + b|1\rangle$ with the required condition that $|\alpha|^2 + |b|^2 = 1$. These coefficients are considered real. The duplicated copy of this state is the 2-qubit state

$$|\varphi^2\rangle \triangleq |\varphi\rangle|\varphi\rangle = a^2|00\rangle + b^2|11\rangle + ab|01\rangle + ab|10\rangle. \tag{23}$$

When applying the CNOT operator (X) with control qubit $|\varphi\rangle$ and controlling (target) state $|0\rangle$, the result is the 2-qubit state

$$|\psi\rangle \triangleq X[|\psi\rangle, |0\rangle] = X[a|0\rangle + b|1\rangle, |0\rangle] = a|00\rangle + b|11\rangle, \tag{24}$$

as it is illustrated in Fig. 1.

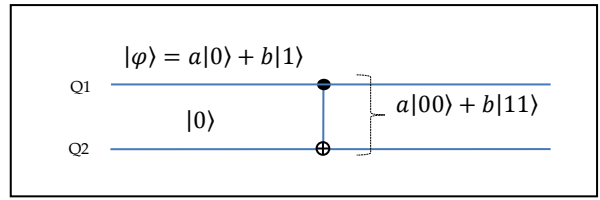

**Fig. 1** The circuit with the CNOT operation.

This operation changes the 2nd qubit state $|0\rangle$ to $|1\rangle$, when the control qubit is $|1\rangle$, i.e., $X[|1\rangle, |0\rangle] = |1\rangle|1\rangle = |11\rangle$. Except for the cases when $a = 0$ and $b = 0$, the states $|\varphi^2\rangle$ and $|\psi\rangle$ are different. Thus, the qubit in its general state is not copying by this circuit with the CNOT gate.

Now, we apply the second CNOT operator with a new control qubit in the state $|\phi\rangle = (|0\rangle + |1\rangle)/\sqrt{2}$, which can be obtained, by using the Walsh-Hadamard gate $H$ on the basis state $|0\rangle$,

$$H|0\rangle = \frac{1}{\sqrt{2}}\begin{bmatrix} 1 & 1 \\ 1 & -1 \end{bmatrix}|0\rangle = \frac{|0\rangle + |1\rangle}{\sqrt{2}}.$$

The target qubit is the second qubit of the 2-qubit state $|\psi\rangle = a|00\rangle + b|11\rangle$. The result of this operation is the following superposition of 3-qubit (without coefficient $1/\sqrt{2}$):

$$|\chi\rangle = X[|\phi\rangle, |\psi\rangle]_2 = X[|0\rangle + |1\rangle, a|00\rangle + b|11\rangle]_2 = X[|0\rangle, a|00\rangle + b|11\rangle]_2 + X[|1\rangle, a|00\rangle + b|11\rangle]_2$$
$$= a|000\rangle + b|011\rangle + a|101\rangle + b|110\rangle. \tag{25}$$

We consider the 3-qubit permutation (1,5)(2,6), for which we will use the gate shown in Fig. 2 in part (a) and call it the 2-XOR operator. This permutation is

$$P_{(1,5),(2,6)}: (0,1,2,3,4,5,6,7) \rightarrow (0,5,6,3,4,1,2,7).$$

The logic element in this figure is not a Toffoli gate over 3-qubit state, which performs the permutation (6,7). The circuit representation of the Toffoli gate is shown in part (b).

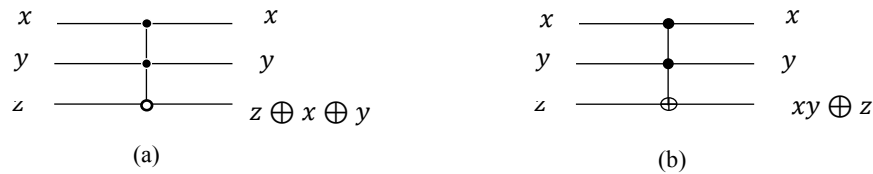

**Fig. 2** Circuit representation of the 3-qubit permutations (a) (1,5)(2,6) and (b) (6,7).

Applying this operator on 3-qubit in superposition of Eq. 25, we obtain the following state:

$$|\xi\rangle = \underbrace{a|000\rangle + b|010\rangle}_{} + \underbrace{a|100\rangle + b|110\rangle}_{} = \underbrace{a|00\rangle + b|01\rangle}_{}|0\rangle + \underbrace{a|10\rangle + b|11\rangle}_{}|0\rangle.$$

It is a superposition of the first four basis states $|0\rangle$, $|1\rangle$, $|2\rangle$, and $|3\rangle$. Using the normalization coefficient $1/\sqrt{2}$, we obtain

$$|\xi\rangle = \frac{a|00\rangle + b|01\rangle}{\sqrt{2}}|0\rangle + \frac{a|10\rangle + b|11\rangle}{\sqrt{2}}|0\rangle, \qquad (26)$$

The abstract circuit for calculating this 3-qubit $|\xi\rangle$ is given in Fig. 3.

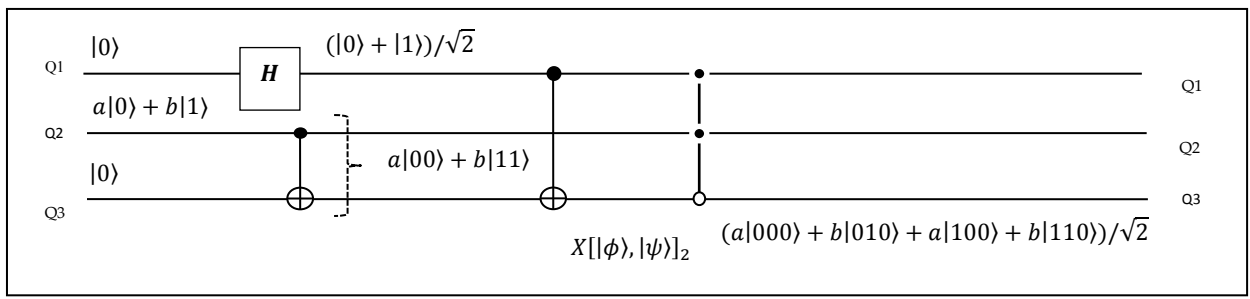

**Fig. 3** The 3-qubit circuit with two CNOT and 2-XOR operations.

The same diagram in compact form is shown in Fig. 4.

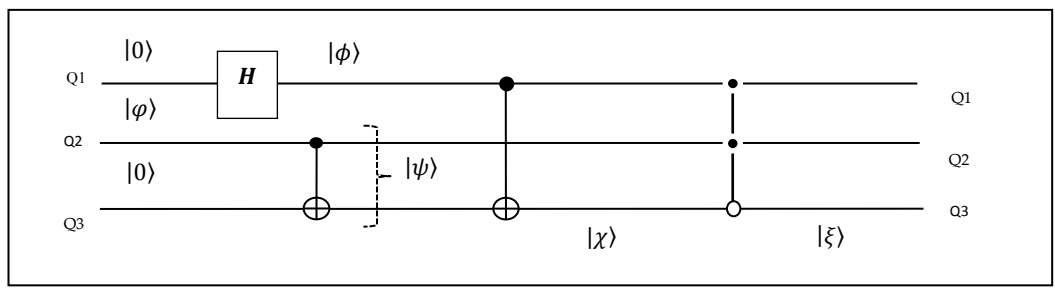

**Fig. 4** The 3-qubit circuit with two CNOT operations and 2-XOR operations.

Now, we consider the doubled qubits $|\varphi\rangle^2$. When measuring the first qubit of $|\varphi\rangle^2$, we obtain with the probability $a^2$ the state $|(\varphi^2)_0\rangle = a|00\rangle + b|01\rangle$, and the state $|(\varphi^2)_1\rangle = a|10\rangle + b|11\rangle$ with probability $b^2$. As follows from Eq. 26, the 3-qubit superposition $|\xi\rangle$ can be written as

$$|\xi\rangle = |(\varphi^2)_0\rangle|0\rangle + |(\varphi^2)_1\rangle|0\rangle = \frac{|(\varphi^2)_0\rangle + |(\varphi^2)_1\rangle}{\sqrt{2}}|0\rangle. \tag{27}$$

Measuring the first qubit of $|\xi\rangle$, we obtain with the same probability 0.5 the following two states. If the measured qubit is 0, the state will be

$$|\xi\rangle_0 = a|000\rangle + b|010\rangle = (a|00\rangle + b|01\rangle)|0\rangle = |(\varphi^2)_0\rangle|0\rangle. \tag{28}$$

If the measured first qubit is 1, then the state will be

$$|\xi\rangle_1 = a|100\rangle + b|110\rangle = (a|10\rangle + b|11\rangle)|0\rangle = |(\varphi^2)_1\rangle|0\rangle. \tag{29}$$

Thus, after measuring the first qubit in the 3-qubit state $|\xi\rangle$, in the first two qubits of the measurement we obtain one of the states of the doubled qubits $|\varphi\rangle^2$, namely

$$|(\varphi^2)_0\rangle = a|00\rangle + b|01\rangle \quad \text{or} \quad |(\varphi^2)_1\rangle = a|10\rangle + b|11\rangle.$$

The last qubit of both measurements is 0. The full circuit of processing the given qubit $|\varphi\rangle$ and measuring the doubled qubits nested in the 3-qubit state is shown in Fig. 5. The parameter of measurement $M = 0$ or 1 when the measured first qubit is 0 or 1, respectively.

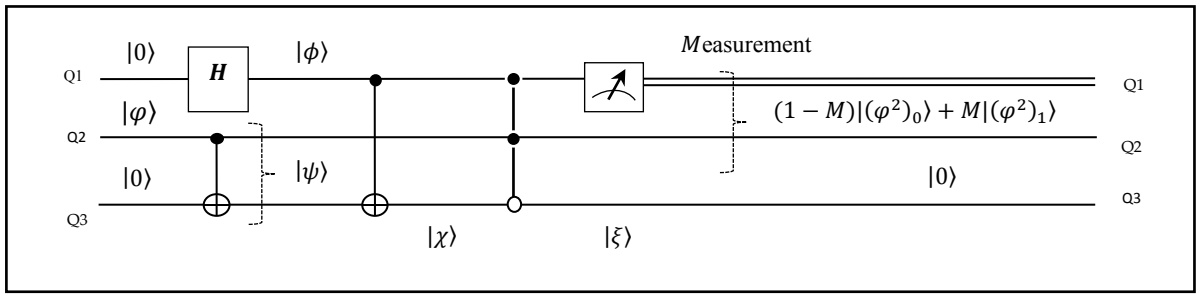

**Fig. 5** The 3-qubit circuit with measurement of the duplicated qubit.

Thus, this circuit shows that the doubled qubits can be nested in the 3-qubit state, namely in the first two qubits of this state. The output of this scheme is a kind of shell containing doubled qubits, from where they can be measured.

**Algorithm of nesting and measuring the doubled qubits in the 3-qubit state:**
1. $|\psi\rangle = X(|\varphi\rangle \oplus |0\rangle)$.
2. $|\phi\rangle = H|0\rangle$.
3. $|\chi\rangle = X[|\phi\rangle, |\psi\rangle]_2$.
4. $|\xi\rangle = P_{(1,5),(2,6)}|\chi\rangle$.
5. Measurement: $|\xi\rangle \rightarrow \{(1-M)|(\varphi^2)_0\rangle + M|(\varphi^2)_1\rangle, |0\rangle\}$.
6. The doubled qubits are described by the first two qubits of the measured 3-qubit,

$$|\varphi\rangle^2 \approx |\xi\rangle_{1,2} = (1-M)\left[a|00\rangle + b|01\rangle\right] + M\left[a|10\rangle + b|11\rangle\right].$$

These measured 2-qubit superpositions carry information of the original qubit $|\varphi\rangle$;

$$a|00\rangle + b|01\rangle = |0\rangle|\varphi\rangle \quad \text{and} \quad a|10\rangle + b|11\rangle = |1\rangle|\varphi\rangle.$$

## Conclusion

The main challenges in quantum computing are not only in developing algorithms and quantum circuits. The accurate measurement of the calculated multi-qubit state is also a difficult task to be solved. Unlike the calculation in the traditional computer, in quantum computing, many measurements are required, i.e., the circuit should be run repeatedly. The quantum circuits to get only the copy of the qubit, i.e., the doubled qubits, are unknown, namely, such circuits do not exist, according to what is said in the current literature. The examples when a unitary transform can be used for transforming two different qubits, but in orthogonal states, are described in this paper. In general, the DsiHT solves this problem easily. In the second part of the paper, we present a quantum circuit with CNOT operations, Hadamard gate, permutation and measurement, which shows the doubled qubits in two qubits of the calculated 3-qubit state. In other words, we have shown that there exist quantum schemes that allow us to measure doubled qubits. As stated in the introduction, in this work we presented our vision of transforming and copying qubits.

### REFERENCES

[1] W.K. Wootters, W.H. Zurek, A single quantum cannot be cloned, *Nature*, vol. 299, 802-803 (1982).
[2] D. Dieks, Communication by EPR Devices, *Phys. Lett.*, vol. 92A, 271-272 (November 1982).
[3] HR. Li, MX. Luo, H. Lai, Generalized quantum no-go theorems of pure states, *Quantum Information Processing* 17 (7), 168, p. 18 (2018).
[4] M. Nielsen, I. Chuang, Quantum Computation and Quantum Information, 2nd Ed., New York, *Cambridge UP* (2001).
[5] E. Strubell, An Introduction to Quantum Algorithms, *Springer*, vol. COS498 (2011).
[6] A. Peres, Quantum Theory: Concepts and Methods, *Kluwer Academic Publishers Dordrecht*, The Netherlands (1995).
[7] C.H. Bennett et al., Purification of noisy entanglement and faithful teleportation via noisy channels, *Phys. Rev. Lett.* 76, 722 (1996).
[8] XQ. Xiao, JM. Liu, Scheme for assisted cloning an unknown arbitrary three-qubit state, *Quantum Information Processing* 10, 567–574 (2011).
[9] V. Buzǐek1, M. Hillery, P.L. Knight, Flocks of quantum clones: Multiple copying of qubits, *Fortschr. Phys.* 46, 4-5, 521-533 (1998).
[10] R. Jozsa, A stronger no-cloning theorem, (2002) arXiv:quant-ph/0204153v2
[11] V. Buzek, M. Hillery, Quantum copying: Beyond the no-cloning theorem, *Phys. Rev.* A 54, 1844 (1996).
[12] L. Duan, G. Guo, A probabilistic cloning machine for replicating two non-orthogonal states, *Phys. Lett.* A 243, 261 (1998).
[13] Kar, G., Rahaman, R.: Local cloning of multipartite entangled states. *Quantum Information Processing* 11, 711–727 (2012).
[14] C. Bennett, G. Brassard, Quantum cryptography: Public key distribution and coin tossing, in Proc. of the IEEE Int. Conf. on Computers, Systems, and Signal Processing, Bangalore, 175-179 (1984).
[15] V. Scarani, S. Iblisdir, N. Gisin, Quantum cloning, (2005) https://arxiv.org/pdf/quant-ph/0511088.pdf
[16] I.B. Djordjevic, Quantum circuits and quantum information processing fundamentals, in Quantum Information Processing and Quantum Error Correction, 91-117 (2012).
[17] A.M. Grigoryan, M.M. Grigoryan, Nonlinear approach of construction of fast unitary transforms, in the Proceedings of the 40th Annual Conference on Information Sciences and Systems (CISS 2006), Princeton Univ., 1073-1078, Princeton (March 22-24, 2006).
[18] A.M. Grigoryan, M.M. Grigoryan, Brief Notes in Advanced DSP: Fourier Analysis with MATLAB, *CRC Press Taylor and Francis Group* (2009).
[19] A.M. Grigoryan, S.S. Agaian, Image processing contrast enhancement, *Wiley Encyclopedia of Electrical and Electronics Engineering*, p. 22 (2017).
[20] A.M. Grigoryan, New method of Givens rotations for triangularization of square matrices, *Journal of Advances in Linear Algebra & Matrix Theory, ALAMT*, vol. 4 (2), 65-78 (2014).
[21] A.M. Grigoryan, Effective methods of QR-decompositions of square complex matrices by fast discrete signal-induced heap transforms, p. 19 (May 10, 2021) arXiv:2105.04089